\documentclass[aps,prd,superscriptaddress,nofootinbib,amsmath,amsfonts,preprintnumbers,notitlepage,10pt,english]{revtex4}
\usepackage{epsf}
\usepackage{amsmath}
\usepackage{graphics}
\usepackage{amsfonts}
\usepackage{amssymb}
\usepackage{latexsym}
\usepackage{color}
\usepackage[dvips]{graphicx}
\usepackage{babel}

\makeatother

\begin{document}
\preprint{RESCEU-34/11}

\title{Evolution of FLRW spacetime after the birth of a cosmic string}

\author{Matthew Lake}
\affiliation{Research Center for the Early Universe (RESCEU),
Graduate School of Science, The University of Tokyo, Tokyo 113-0033, Japan}

\author{Teruaki Suyama}
\affiliation{Research Center for the Early Universe (RESCEU),
Graduate School of Science, The University of Tokyo, Tokyo 113-0033, Japan}

\begin{abstract}
We consider the evolution of an initially FLRW universe after the formation of a long, straight, cosmic string with arbitrary tension and mass per unit length. The birth of the string sources scalar and tensor-type perturbations in the background metric and both density and velocity perturbations in the background fluid, which compensate for the string mass and maintain energy conservation. The former generate the deficit angle within the light cone of the string and a gravitational shock front at the cosmological horizon, whereas the latter are confined within the sound cone. We study the properties of the metric within each region of the resulting spacetime and give the explicit coordinate transformations which demonstrate non-violation of causality. This paper generalizes the work of previous studies for the Nambu-Goto string. 
\end{abstract}

\date{\today}

\maketitle

\section{Introduction}
Cosmic strings are linear concentrations of energy which may have formed during symmetry-breaking phase transitions in the early Universe \cite{Nielsen_Olesen, hep-ph/9411342, VS1994,hep-th/0508135v2, Preskill}. 
Though the string width is determined by the inverse of the symmetry-breaking energy scale, this is small compared to cosmological distances, and they may be approximated as one-dimensional objects for many purposes \cite{Goto1,Anderson2003}. 
Specifically, strings may have been produced at the epoch of electro-weak symmetry-breaking \cite{Nambu1}, or the GUT scale, but their formation is also a generic feature of the phase transitions inherent in many extensions of the standard model,  (c.f. \cite{hep-ph/0308134} and references therein).
In field theory they are a type of topological defect, analogous to the magnetic flux tubes and other vortex-type defects created in certain condensed matter systems \cite{Abrikosov1, Zurek1, Zurek2, Zurek3,  Bowick1, Williams1, Hendry1, Chuang1, Annett1}, and are produced via the Kibble mechanism \cite{ICTP/75/5},
if the vacuum manifold ($\mathcal{M}$) possesses a nontrivial first homotopy group (e.g. $\pi_1(\mathcal{M})=\mathbb{Z}$, in which each element, a nonzero integer, corresponds to an allowed winding number for the string vortex cross-section). In recent years, the formation of ``cosmic", i.e. horizon-sized, fundamental strings ($F$-strings) and one-dimensional $D$-branes ($D$-strings) has also been extensively studied in string theory \cite{Witten:1985, Polchinski_Intro, CMP_FD1,CMP_FD2,0911.1345v3, 0811.1277v1, hep-th/0505050v1, astro-ph/0410073v2, Sakellariadou:2009, Rajantie:2007, Copeland:2005}, particularly in the context of brane inflation (see \cite{Cline1, Gasperini1, Tye1, Carroll:TASI, Quevedo:Lectures, Danielsson1} for reviews), in which such defects can be copiously produced \cite{Sarangi1, Jones_etal1, Pogosian_Obs1}.
However, regardless of the precise details of individual models, the main phenomenological, and observationally relevant, parameters that characterize the string are the energy of the per unit length,
which we denote $U$, and the tension, $\mu = qU$.
\footnote{This is not strictly true in the case of models containing compact extra-dimensions in which the intercommuting probability for strings which ``cross" in the noncompact space can be much less than unity ($P<<1$). In this case the dynamics of string networks and their implications for cosmological observations can be substantially altered \cite{Sarangi1,Jones_etal1,Pogosian_Obs1,Avgoustidis:2004,Copeland:2005}. 
Extra-dimensional scenarios, especially those motivated by, or directly embedded into, string theory also allow for the formation of cosmic ``necklaces" of various kinds, which introduce significant alterations to the network dynamics along with additional phenomenological parameters, such as the bead mass \cite{Leblond:2007, Dasgupta1, Martins1, Matsuda2, Lake1, Lake3}. The analogues of a superstring necklaces in field theory are series of monopoles connected by string-segments, which may also be formed in some symmetry-breaking models \cite{Siemens1, Blanco-Pillado1}. The authors hope to address the important question of the inborn metric for a cosmic necklace in a future article.}
The Nambu-Goto (NG) action represents the simplest model and enforcing Lorentz symmetry along the string
corresponds to setting $q=1$.
If the NG string is wiggly at microscopic scales, it may be approximated macroscopically by a simpler embedding with an effective mass per unit length which differs from the intrinsic tension. Thus in the coarse-grained limit, small-scale structure, even on pure NG strings, can lead to a value of $q$ different from unity \cite{vilenkin,carter}.
For current carrying strings, the mass-density and tension may differ, as originally shown by Witten \cite{witten,FPRINT-92-39}, so that $q \neq 1$ in general.
\\ \indent
The gravitational field around a static, straight string exhibits an interesting feature
not seen in fields surrounding spherically-symmetric distributions of matter. 
In a perturbative analysis of a string in the wire approximation (to first order in $G\mu$), Vilenkin \cite{167218} showed that test particles within the string light-cone experience a gravitational acceleration
along the radial direction equal to $2GU(q-1)/r$, where $r$ is a distance from the string,
and that the surrounding space has a conical structure with deficit angle $4\pi (1+q)GU$. For the NG string therefore, the gravitational force vanishes and the spacetime becomes
locally Minkowski. These results were found to hold in a nonperturbative analysis (i.e. to all orders in $G\mu$), independently, by Gott \cite{Gott1} and Hiscock \cite{Hiscock1}, even for a string of finite width. 
One observational consequence is the formation of double images from bright sources lying behind the string in relation to an observer. 
The images have exactly the same shape and brightness and for NG strings their angular separation on the sky is of the order of the deficit angle, though for $q>>1$ it may be substantially smaller \cite{Pe94,Uz01}. This feature has been used to search for long strings  
\cite{Gott1, Vilenkin_lensing1,Hogan_lensing1,Shlaer_lensing1,Huterer_lensing1,Hindmarsh_lensing1, Dyda_lensing1,Gasperini_lensing1, Morganson_lensing1, deLaix_lensing1} and loops \cite{Cowie_lensing1, deLaix_lensing2, Mack_lensing1} via astronomical observations. Bounds on phenomenological parameters such as the string tension and number density have also been obtained using the predicted effects of strings on CMB anisotropy \cite{Zeldovich_CMB1, Kaiser_CMB1,Stebbins_CMB1,Gangui_CMB1,Allen_CMB1,Perivolaropoulos_CMB1,Benabed_CMB1,Landriau_CMB1, Landriau_CMB2, Pogosian_CMB1, Wyman_CMB1, Battye_CMB1, Battye_CMB2, Contaldi_CMB1, Bevis_CMB1, Bevis_CMB2, Bevis_CMB3, Bevis_CMB4,Suyama_CMB1,Suyama_CMB2}, 
the expected emission of high-energy rays from cusps and kinks \cite{Bhattacharjee_CosRay1, Sigl_CosRay1, Berezinsky_CosRay1, Wichoski_CosRay1} and possible signatures in the 21cm line \cite{Khatri_21cm}.  Interestingly, it has recently been proposed that cusp-emission from superconducting strings may be responsible for anomalous gamma-ray burst observations \cite{Gr09,Ta09,Sa09} and there remains ongoing controversy on that point (c.f. \cite{Ba87,Pa88,Br93,Pl94,Be01,Be04,Ch10,Wa11,Ch11} and references therein). With the advent of the Planck experiment \cite{Planck}, it is also hoped that the polarization of CMB $B$-modes caused by string networks in the early universe could confirm, or rule out, the existence of GUT-scale strings in the near future \cite{Seljak_Pol1,Pogosian_Pol1,Pogosian_Pol2,Bevis_Pol1,Garcia_Pol1}. Finally, gravitational radiation from strings has been extensively studied \cite{Vachaspati_GravRad1,Hindmarsh_GravRad1,Darmour_GravRad1,Darmour_GravRad2,Darmour_GravRad3}
and future gravitational wave detections from experiments such as LISA/NGO \cite{LISA/NGO} or LIGO \cite{LIGO} may place even stronger bounds on string parameters. 
\\ \indent
It is interesting to consider how the conical structure of the spacetime surrounding the string evolves from the initial Friedmann-Lema\^\i tre-Robertson-Walker (FLRW) 
background, which exists before the symmetry-breaking epoch, and a detailed analysis of this problem was originally given by Magueijo \cite{Magueijo:1992tt,mag-unpublish} for the case of an NG string. He introduced a phenomenological string energy-momentum tensor which is localized at the
origin and appears only after the symmetry-breaking time.
The generation of string mass-energy is compensated by a decrease in the mass-energy of the perfect fluid 
that dominates the Universe, so that conservation of energy and momentum are not violated. \footnote{In previous analyses
this was not always the case, though particle production by cosmic strings was studied in detail in the following sources, \cite{Parker:1987qx,Sa88,Da88}.}
Inclusion of the background fluid is therefore compulsory and the problem of string generation must be considered on an (initially) FLRW background.
In his analysis, Magueijo obtained solutions of the linearized Einstein equations on the FLRW background metric by assuming the
dimensionless parameter $GU$ to be small and constant. This is consistent with known observational bounds on the string tension in various cosmological models, which give an upper limit of $GU \leq \mathcal{O}(10^{-6})$ for field-theoretic strings \cite{VS1994, Bevis_CMB1,Bevis_CMB2,Bevis_CMB3,Bevis_CMB4} and $10^{-11} \leq GU \leq 10^{-6}$ for the warped tension of cosmic $F$/$D$-strings \cite{Jones_etal1,Pogosian_Obs1} (see also \cite{Hindmarsh_rev} for a review of observational constraints up to 1990).
He showed that, at late times and in the region close to the string, the spacetime takes the form of that surrounding an eternal string having deficit angle $8\pi GU$ \cite{167218,Gott1,Hiscock1} and that a gravitational shock propagates along the lightcone, with the spacetime outside the horizon remaining unperturbed so that the FLRW metric remains valid, though detailed calculations confirming this picture remain unpublished \cite{mag-unpublish}.
\\ \indent
In this paper, we reconsider this issue and also extend the original analysis by considering the more general case of $q \ne 1$. This allows us to include the physically interesting class of ``conical" spacetimes which are not locally flat and to consider the gravitational acceleration of test particles in the vicinity of a noneternal string.
We follow the same approach taken in \cite{Magueijo:1992tt} to evaluate the metric perturbations induced
by string formation and find that the general picture remains qualitatively similar. However, as shown in detail in Sec.~IV, our solutions do not exactly match those obtained previously, even for $q=1$.
We believe that this is because of the choice of boundary condition for the Poisson equation governing 
the gravitational potential given in \cite{Magueijo:1992tt} is not consistent with trace part of the Einstein equations.
We also provide concrete coordinate transformations both for inside and outside
the sound horizon of the perfect fluid, as well as in the region between the sound horizon and the edge of the light cone, which result in an FLRW metric beyond the causal horizon and the metric for an eternal string deep inside the sound horizon, respectively, in accordance with previous results \cite{167218,Gott1,Hiscock1,Magueijo:1992tt,mag-unpublish}. 
\\ \indent
The structure of the paper is as follows: In Sec.~II we define the perturbation variables and decompose them into
scalar and tensor components.
In Sec.~III we derive the master equations for each type, while analytic solutions are given in Sec.~IV together with a comparison of the corresponding solutions obtained in \cite{Magueijo:1992tt}.
Section V is devoted to studying the properties of the resulting class of spacetimes and a brief summary of our results and discussion of the prospects for future work is given in the conclusion, Sec.~VI.
\section{Definition of the Perturbation variables}
\subsection{Background dynamics}
We assume that the background spacetime is an FLRW universe with metric
\begin{equation}
ds^2=a(\eta)^2 \left( -d\eta^2+ \gamma_{ij} dx^i dx^j \right),
\end{equation}
where $\gamma_{ij}={\rm diag}(1,1,r^2)$ is the spatial part in cylindrical coordinates and $\eta$ is the conformal time coordinate, defined via $a(\eta)d\eta = dt$.
We also assume that the universe expands due to the presence of a perfect fluid with energy-momentum tensor
\begin{equation}
{}_m T^\mu_{~\nu}=(\rho+P) u^\mu u_\nu+P \delta^\mu_{~\nu}.
\end{equation}
where $\rho$ is the energy density, $P$ is the pressure and, from the normalization condition $u^\mu u_\mu=-1$, we have that
\begin{equation}
u^\mu=\left( \frac{1}{a},0,0,0 \right),~~~~~u_\mu=(-a,0,0,0).
\end{equation}
The Friedmann equation and continuity equation are given, respectively, by
\begin{equation}
{\cal H}^2=\frac{8\pi G}{3}\rho a^2,~~~{\dot \rho}+3(1+c_s^2){\cal H}\rho=0,
\end{equation}
where ${\dot {}}$ indicates the derivative with respect to conformal time and from these two equations, we find that
\begin{equation}
{\cal H}=\frac{2}{(1+3c_s^2) \eta},~~~{\dot {\cal H}}=-\frac{1}{2}(1+3c_s^2) {\cal H}^2.
\end{equation}
\subsection{Metric perturbation}
The scalar part of the perturbed metric may be diagonalized using gauge degrees of freedom and the full metric can be written in the form,
\begin{equation}
ds^2=a(\eta)^2 \bigg[ -(1+2 \Phi) d\eta^2+(1+2\Psi) \gamma_{ij} dx^i dx^j+h_{ij} dx^i dx^j \bigg].
\end{equation}
The gauge invariant tensor-type perturbations $h_{ij}$ satisfy the transverse and
traceless conditions,
\begin{equation}
D_i h^i_{~j}=h^i_{~i}=0,
\end{equation}
where $D_i$ is the covariant derivative defined for the three-dimensional metric $\gamma_{ij}$.
We do not consider vector-type perturbations since they are not generated by the situation we are interested in (see Subsec. \ref{subsec-C}).
The nonzero components of the perturbed Einstein tensor for the scalar-type perturbations are given by
\begin{eqnarray}
&&\delta G^0_{~0}=\frac{2}{a^2} \left( 3 {\cal H}^2 \Phi+D^2 \Psi -3{\cal H} {\dot \Psi} \right), \\
&&\delta G^0_{~i}=\frac{2}{a^2} D_i \left( {\dot \Psi}-{\cal H}\Phi \right), \\
&&\delta G^i_{~j}=\frac{1}{a^2} \left( D^i D_j-\frac{1}{3} \delta^i_{~j} D^2 \right) \Sigma_g+\frac{1}{3a^2} \delta^i_{~j} \Pi_g, 
\end{eqnarray}
where we define $\Sigma_g$ and $\Pi_g$ according to
\begin{equation}
\Sigma_g=-\Phi-\Psi,~~~~~\Pi_g=2D^2 (\Phi+\Psi)-6 \left( {\cal H}^2-2 \frac{\ddot a}{a} \right) \Phi +6{\cal H} \left( {\dot \Phi}-2{\dot \Psi} \right)-6 {\ddot \Psi}.
\end{equation}
For the tensor-type perturbations, it transpires that only diagonal components are excited
by the string energy-momentum tensor, so that we may set $h^r_{~z}=h^r_{~\theta}=h^z_{~\theta}=0$ without loss of generality.
The tranverse and traceless conditions also allow us to write $h^z_{~z}$ and $h^\theta_{~\theta}$ solely in terms of $h^r_{~r}$, so that $h^r_{~r}$ is the only independent component;
\begin{equation}
h^\theta_{~\theta}=h^r_{~r}+r \frac{\partial}{\partial r} h^r_{~r},~~~~~h^z_{~z}=-2 h^r_{~r}-r \frac{\partial}{\partial r} h^r_{~r}.
\end{equation}
Rewriting $h^r_{~r} = F$, 
the relevant component of the perturbed Einstein tensor is the  $r-r$ component, given by
\begin{equation}
\delta G^r_{~r}=\frac{1}{2 a^2} \left( {\ddot F}+2{\cal H}{\dot F}- F''-\frac{3}{r} F' \right).
\end{equation}
\subsection{Energy-momentum tensor of a cosmic string}\label{subsec-C}
As mentioned previously, in field theories, the string width is roughly equal to the inverse of the symmetry-breaking energy scale which, in realistic models, is likely to be above the electroweak scale \cite{hep-ph/0308134}. For cosmic superstrings it is expected to be of the order of the fundamental string scale which, although unknown, may be as small as the Planck length \cite{Polchinski_Intro}. Since both of these length-scales are typically much smaller than the Hubble horizon, the finite width of the string does not play a cosmologically significant role as long as we are concerned with purely gravitational effects. 
The finite duration of the string formation as a result of the phase transition is also much shorter 
than the Hubble time and is not significant on cosmological timescales.
As considering these effects would complicate the analysis of our problem immensely, we avoid such complexity by assuming that the string width is zero (adopting the wire approximation), and that string formation occurs instantly.
\\ \indent
With this approximation, the energy-momentum tensor of the cosmic string can be written as 
\begin{equation} \label{eq:TCS1}
{}_{(\rm CS)}T^\mu_{~\nu}=-\frac{U \delta (r)}{2\pi a^2 r} \Theta (\eta-\eta_0) {\rm diag}(1,0,q,0),
\end{equation}
where $U$ is the energy density of string per unit length and $\mu = qU$ is the tension.
\\ \indent
For current carrying strings and coarse-grained strings that are wiggling on microscopic scales, 
$q$ generically takes a value different from unity.
Here, we do not specify microscopic model of the string and will leave $q$ as a free parameter. 
\\ \indent
We begin by decomposing ${}_{(\rm CS)}T_{ij}$ into scalar, vector and tensor-type perturbations
\begin{equation} 
{}_{(\rm CS)}T_{ij}=\left( D_i D_j-\frac{1}{3} \gamma_{ij} D^2 \right) \Sigma_{(\rm CS)}+\frac{1}{3} \gamma_{ij} \Pi_{(\rm CS)}+D_i v_j+D_j v_i+H_{ij}, \label{cs-decompose}
\end{equation}
where ${}_{(\rm CS)}T_{ij}=g_{ik} {}_{(\rm CS)}T^k_{~j}=a^2 \gamma_{ik} {}_{(\rm CS)}T^k_{~j},~D^i v_i=0$ and  $H_{ij}=\gamma^{ij}H_{ij}=0$.
Comparing Eqs. (\ref{eq:TCS1}) and (\ref{cs-decompose}) then implies
\begin{eqnarray} \label{eq:basic_pert_eqns}
&&\Pi_{(\rm CS)}=-\frac{qU \delta (r)}{2\pi r} \Theta (\eta-\eta_0), \\
&&\Sigma_{(\rm CS)}=\frac{qU}{4\pi} \log r ~\Theta (\eta-\eta_0), \\
&&v_i=0, \\
&&H_{rr}=\frac{qU}{4\pi r^2} \Theta (\eta-\eta_0), \\
&&H_{rz}=0,\\
&&H_{r\theta}=0, \\
&&H_{zz}=-\frac{qU \delta (r)}{4\pi r}\Theta (\eta-\eta_0), \\
&&H_{z\theta}=0,\\
&&H_{\theta \theta}=\frac{qU r\delta (r)}{4\pi} \Theta (\eta-\eta_0)-\frac{qU}{4\pi}\Theta (r) \Theta (\eta-\eta_0).
\end{eqnarray}
Since ${}_{(\rm CS)} T_{\mu \nu}$ is itself a first order quantity, ${}_{(\rm CS)}T^i_j$ is simply given by
\begin{equation}
{}_{(\rm CS)}T^i_j=g^{i\mu} {}_{(\rm CS)} T_{\mu j}=\frac{1}{a^2} \bigg[ \left( D^i D_j-\frac{1}{3} \delta^i_{~j} D^2 \right) \Sigma_{(\rm CS)}+\frac{1}{3} \delta^i_{~j} \Pi_{(\rm CS)}+H^i_{~j} \bigg].
\end{equation}
where $H^i_{~j} = \gamma^{ik} H_{kj}$.
\subsection{Energy-momentum tensor of a perfect fluid}
We assume that the perfect fluid obeys the equation of state given by $P=c_s^2 \rho$, where $c_s$ is the sound speed.
We then write the velocity perturbation as $u_i = a D_i v$ so that the perturbed four-velocity is given by
\begin{equation}
u^\mu=\left( \frac{(1-\Phi)}{a}, \frac{1}{a} D^i v \right),~~~u_\mu=\left( -a (1+\Phi),a D_i v \right).
\end{equation}
The nonzero components of the perturbed energy-momentum tensor  are then
\begin{eqnarray}
&&\delta T^0_{~0}=-\delta \rho_m, \\
&&\delta T^0_{~i}=(\rho+P) D_i v, \\
&&\delta T^i_{~j}=c_s^2 \delta \rho_m \delta^i_{~j}.
\end{eqnarray}
\section{Perturbation equations}
\subsection{Scalar-type perturbations}
Six equations may be derived for the scalar-type variables, two from the conservation laws
of the energy-momentum tensor and four from the Einstein equations. 
These constitute the basic equations that govern the evolution of the density 
contrast ($\delta_m=\delta \rho/\rho$) and velocity perturbation ($v$) of the perfect fluid, the gravitational potential ($\Phi$) and the curvature perturbation ($\Psi$).
However, not all these equations are independent as they are related through Bianchi identities and it is possible to derive a closed differential equation for each variable.
Before doing this, let us first take the following ansatz;
\begin{equation}
\delta_m={\tilde \delta_m} \Theta (\eta-\eta_0),~~~v={\tilde v} \Theta (\eta-\eta_0),~~~
\Phi={\tilde \Phi} \Theta (\eta-\eta_0),~~~\Psi={\tilde \Psi} \Theta (\eta-\eta_0). \label{scalar-ansatz}
\end{equation}
This is reasonable since nothing must be excited before the phase-transition epoch, $\eta=\eta_0$. The basic perturbation equations 
can be written in the following form;
\begin{eqnarray}
&&{\ddot {\tilde \Psi}}+3(1+c_s^2) {\cal H} {\dot {\tilde \Psi}}-c_s^2 D^2 {\tilde \Psi}=\frac{2c_s^2 GU \delta (r)}{r}, \label{master-psi}\\
&&{\tilde \Phi}+{\tilde \Psi}=-2qGU \log r, \label{trace-part}\\
&&D^2 {\tilde \Phi}=2(1-q) GU \frac{\delta (r)}{r}+\frac{3{\cal H}^2}{2} {\tilde \Delta_m}, \label{poisson-eq} \\
&&{\dot {\tilde \Psi}}-{\cal H} {\tilde \Phi}=\frac{3}{2} (1+c_s^2) {\cal H}^2 v, \label{time-space}
\end{eqnarray}
where ${\tilde \Delta_m}$ is defined by
\begin{equation}
{\tilde \Delta_m} \equiv {\tilde \delta_m}-3(1+c_s^2) {\cal H} {\tilde v}.
\end{equation}
Equation (\ref{master-psi}) is a closed differential equation for ${\tilde \Psi}$.
Once the time evolution of ${\tilde \Psi}$ is determined, the other variables 
${\tilde \Phi},~{\tilde \Delta_m}$ and ${\tilde v}$ are detemined uniquely
by Eqs.~(\ref{trace-part})-(\ref{time-space}).
The initial conditions for ${\tilde \Psi}$ are given by
\begin{eqnarray}
{\tilde \Psi}(\eta_0,r)=0,~~~{\dot {\tilde \Psi}}(\eta_0,r)=-2qGU {\cal H} \log r. \label{ini-scalar}
\end{eqnarray}
and are obtained by requiring that the terms containing a delta-function $\delta (\eta-\eta_0)$,
which appears in the perturbation equations after the change of variables Eq.(\ref{scalar-ansatz}),
should vanish.
\\ \indent
We can simplify the master equation for ${\tilde \Psi}$ (\ref{master-psi}) by introducing a new variable, ${\cal R}$, defined as
\begin{equation} \label{eq:R_defn}
{\cal R}={\tilde \Psi}+2GU \log r.
\end{equation}
The evolution equation for ${\cal R}$ is
\begin{equation}
{\ddot {\cal R}}+3(1+c_s^2) {\cal H} {\dot {\cal R}}-c_s^2 D^2 {\cal R}=0, \label{master-scalar},
\end{equation}
which is free from the source term given by the appearance of the string.
Conveniently, all information regarding the string formation is contained solely in the initial conditions although, physically, this may have been expected since (at least within our current approximation), this produces a ``source" term for the metric perturbations only instantaneously.
The initial conditions for ${\cal R}$ are
\begin{equation}
{\cal R}(\eta_0,r)=2GU \log r,~~~{\dot {\cal R}}(\eta_0,r)=-2qGU {\cal H} \log r. \label{initial-scalar}
\end{equation}
\subsection{Tensor-type perturbations}
As mentioned in Sec. II-B, the relevant evolution equation for the tensor perturbation 
is given by the $r-r$ component of the Einstein equations,
\begin{equation}
{\ddot F}+2{\cal H}{\dot F}- F''-\frac{3}{r} F'=\frac{4qGU}{r^2} \Theta (\eta-\eta_0). \label{tensor-new-eq}
\end{equation}
It may be confirmed that the evolution equations for the other components can be derived from this.
Taking the ansatz $F(\eta,r)={\tilde F}(\eta,r) \Theta (\eta-\eta)$, Eq.~(\ref{tensor-new-eq}) may then be rewritten as
\begin{equation}
{\ddot {\tilde F}}+2{\cal H}{\dot {\tilde F}}-{\tilde F}''-\frac{3}{r} {\tilde F}'=\frac{4qGU}{r^2}, \label{tensor-evolve}
\end{equation}
and the initial conditions are,
\begin{eqnarray}
{\tilde F}(\eta_0,r)={\dot {\tilde F}}(\eta_0,r)=0.
\end{eqnarray}
\section{Solving perturbation equations}
\subsection{Scalar-type perturbations}
Let us try solving the master equation for ${\cal R}$ (\ref{master-scalar}) subject to the initial conditions (\ref{initial-scalar}).
We first perform the Fourier-transformation for ${\cal R}$;
\begin{equation}
{\cal R}(\eta,r)=\int \frac{d^2k}{{(2\pi)}^2} e^{i {\vec k} \cdot {\vec x}} {\cal R}_k (\eta), \label{Fourier-scalar}
\end{equation}
where ${\vec x}=r (\cos \theta,~\sin \theta)$ is two-dimensional vector on the $(r,~\theta)$-plane.
The evolution equation for each Fourier component is
\begin{equation}
{\ddot {\cal R}_k}+\frac{2(\nu+1)}{\eta} {\dot {\cal R}_k}+c_s^2 k^2 {\cal R}_k=0, \label{Fourier-scalar-eq},
\end{equation}
where $\nu$ is defined as $\nu=\frac{2}{1+3c_s^2}$.
In order to give the initial conditions for each ${\cal R}_k$, we must also Fourier-decompose the $\log r$ term in Eq. (\ref{eq:R_defn}).
Rewriting $\log r$ as 
\begin{equation}
\log r =\int \frac{d^2 k}{{(2\pi)}^2} e^{i {\vec k} \cdot {\vec x}} \alpha_k, \label{Fourier-logr}
\end{equation}
where $\alpha_k$ are constants, we find that the identity $D^2 \log r=\frac{\delta (r)}{r}$ yields $\alpha_k=-2\pi/k^2$.
However, direct substitution of this into Eq.~(\ref{Fourier-logr}) results in
a divergence which does not depend on $r$.
To extract $r$-dependent part we must therefore modify the coefficients via the transformation $\alpha_k \rightarrow \tilde{\alpha}_k =-2\pi/k^{2-\epsilon}$, where
$\epsilon$ is a positive infinitesimal.
This choice of $\tilde{\alpha}_k$ yields
\begin{equation}
-\int \frac{d^2 k}{2\pi} \frac{e^{i {\vec k} \cdot {\vec x}}}{k^{2-\epsilon}}=-\frac{1}{\epsilon}+\log r+{\rm const.}+{\cal O}(\epsilon).
\end{equation}
so that the part which is independent of $\epsilon$ gives the desired $\log r$ (the {\rm const.} term
is not important for our purpose).
To find the desired solution for ${\cal R}_k$, we keep $\epsilon$ in the intermediate calculations before
letting $\epsilon \rightarrow 0$ in the final expression, thus picking up terms which are independent of $\epsilon$.
With this prescription, the initial conditions for ${\cal R}_k$ can be written as
\begin{equation}
{\cal R}_k=-\frac{4\pi GU}{k^{2-\epsilon}},~~~~~{\dot {\cal R}_k}=\frac{4\pi q\nu GU}{k^{2-\epsilon} \eta_0}.
\end{equation}
It is then straightforward to derive the solution of Eq.~(\ref{Fourier-scalar-eq}), which is
\begin{equation}
{\cal R}_k (\eta)=\frac{4\pi GU c_s \eta_0^{\nu+1}}{k^{1-\epsilon} \eta^\nu} \left( a_k j_\nu (c_s k \eta)+b_k n_\nu (c_s k\eta) \right),
\end{equation}
where $j_\nu$ and $n_\nu$ are spherical Bessel functions of the first and second kind, respectively, and 
$a_k$ and $b_k$ are given by
\begin{equation}
a_k=-q\nu n_\nu (c_s k\eta_0)+c_s k \eta_0 n_{\nu+1}(c_s k\eta_0),~~~b_k=q\nu j_\nu (c_s k\eta_0)-c_s k \eta_0 j_{\nu+1}(c_s k\eta_0).
\end{equation}
Using this solution with arbitrary values of the equation of state parameter $c_s$, and the string formation time $\eta_0$,
we can obtain the time evolution of ${\cal R}(\eta,r)$ by integrating Eq.~(\ref{Fourier-scalar}) 
over $k$ numerically.
\\ \indent
However, let us consider a situation where an analytic evaluation of the integral (\ref{Fourier-scalar}) may be obtained to good approximation.
We begin by assuming that $\eta_0$ is sufficiently small, such that both $r$ and the sound horizon distance, $c_s \eta$, are much larger $\eta_0$.
This approximation applies to a comoving observer attached to the background FLRW universe 
(and whose position is therefore specified uniquely by the comoving radius $r$), at a time much later than the time of string formation.
We do not assume a magnitude relation between $c_s\eta$ and $r$, that is, $c_s\eta$ 
can be larger or smaller than $r$.
If $c_s \eta >r$ the observer is inside the sound cone, whereas $c_s \eta <r$ means that the observer lies outside the sound horizon.
\\ \indent
Since the contribution to ${\cal R} (\eta,r)$ from the $b_k$  part is suppressed by
positive powers of $\eta_0$, it vanishes in the limit $\eta_0 \to 0$.
We therefore neglect the contribution from $b_k$ and keep only $a_k$ term.
We then Taylor expand the spherical Bessel functions in $c_s k \eta_0$ and keep only the leading order term, so that
\begin{equation}
a_k \simeq \frac{(q\nu-2\nu-1) \Gamma (2\nu+1)}{2^\nu \Gamma (\nu+1)} {(c_s k \eta_0)}^{-\nu-1}.
\end{equation}
With this approximation, ${\cal R}(\eta,r)$ becomes
\begin{equation}
{\cal R}(\eta,r) \simeq 2^{1-\nu} GU (q\nu-2\nu-1) \frac{\Gamma(2\nu+1)}{\Gamma (\nu+1)} {(c_s \eta)}^{-\epsilon} \int_0^\infty dz~z^{\epsilon-\nu-1} J_0 (xz) j_\nu(z),
\end{equation}
where $x$ is defined by $x \equiv r/(c_s \eta)$.
Using the formula for the integral of a product of two Bessel functions given in the appendix,
we find that integration over $z$ yields,
\begin{equation}
\int_0^\infty dz~z^{\epsilon-\nu-1} J_0 (xz) j_\nu(z)=\frac{\sqrt{\pi} 2^{-\nu-2+\epsilon} \Gamma \left( \frac{\epsilon}{2} \right)}{x^\epsilon \Gamma \left( \nu+\frac{3}{2} \right) \Gamma \left( 1-\frac{\epsilon}{2} \right)} F\left( \frac{\epsilon}{2},\frac{\epsilon}{2},\nu+\frac{3}{2};x^{-2} \right)~~~~~{\rm for}~x>1,
\end{equation}
and
\begin{equation}
\int_0^\infty dz~z^{\epsilon-\nu-1} J_0 (xz) j_\nu(z)=\frac{\sqrt{\pi} 2^{-\nu-2+\epsilon} \Gamma \left( \frac{\epsilon}{2} \right)}{\Gamma \left( \nu+\frac{3}{2}-\frac{\epsilon}{2} \right)} F\left( \frac{\epsilon}{2},-\nu-\frac{1}{2}+\frac{\epsilon}{2},1;x^2 \right)~~~~~{\rm for}~x<1,
\end{equation}
where $F(a,b,c;x)$ is Gauss' hypergeometric function.
After Taylor expanding these expressions and picking up terms independent of $\epsilon$ by allowing $\epsilon \rightarrow 0$ (as before), we find that ${\cal R}$ is given by
\begin{equation}
{\cal R}(\eta,r)=\bigg[ -\frac{2(q\nu-2\nu-1)GU}{2\nu+1} \log r+{\rm const.} \bigg] \Theta (x-1)+\frac{(q\nu-2\nu-1)GU}{2\nu+1}\left( -2 \log \eta+ {\cal G}_\nu (x) \right) \Theta (1-x),
\end{equation}
where  we define the function ${\cal G}_\nu (x)$ as
\begin{equation}
{\cal G}_\nu (x) \equiv \frac{\partial}{\partial b} F\left( b,-\frac{1}{2}-\nu,1;x^2 \right) \bigg|_{b=0}.
\end{equation}
For later convenience, we here give an explicit expression for ${\cal G}_\nu$ when $\nu=1$ in terms of elementary functions;
\begin{equation}
{\cal G}_1(x)=2\left( \sqrt{1-x^2}+\frac{1}{3} {(1-x^2)}^{3/2}-\log ( 1+\sqrt{1-x^2})-\frac{4}{3}+\log 2 \right).
\end{equation}
This corresponds to the physically interesting case $c_s^2=1/3$, i.~e.~a radiation-dominated universe.
Having determined ${\cal R}$, we can immediately determine ${\tilde \Psi}$ from  Eq.~(\ref{eq:R_defn});
\begin{equation}
{\tilde \Psi}=-\frac{2q\nu GU}{2\nu+1} \log r ~\Theta(x-1)+GU \bigg[ -2 \log r+\frac{(q\nu-2\nu-1)}{2\nu+1} (-2 \log \eta+{\cal G}_\nu (x) ) \bigg] \Theta (1-x), \label{full-psi}
\end{equation}
followed by ${\tilde \Phi}$ from Eq.~(\ref{trace-part}).
\subsection{Tensor-type perturbations}
By performing the transformation $\psi = r{\tilde F}$, the evolution equation for ${\tilde F}$ (\ref{tensor-evolve}) may be rewritten in terms of the new variable $\psi$ as
\begin{equation}
{\ddot \psi}+\frac{2\nu}{\eta} {\dot \psi}-\psi''-\frac{1}{r} \psi'+\frac{1}{r^2}\psi=\frac{4qGU}{r}.
\end{equation}
A solution of this equation satisfying the initial conditions $\psi (\eta_0,r)={\dot \psi}(\eta_0,r)=0$ is
\begin{equation}
\psi (\eta,r)=\int_0^\infty dk~A_k (\eta) J_1 (kr),
\end{equation}
where $A_k(\eta)$ is a function that obeys the following differential equation,
\begin{equation}
{\ddot A_k}+\frac{2\nu}{\eta}{\dot A_k}+k^2 A_k=4qGU.
\end{equation}
Imposing the initial conditions $A_k(\eta_0)={\dot A_k}(\eta_0)=0$,
we find that $A_k (\eta)$ is given by
\begin{equation}
A_k (\eta)=\frac{4qGU}{k^2} \bigg[ 1-k\eta k \eta_0 {\left( \frac{\eta_0}{\eta} \right)}^\nu \big\{ j_\nu (k\eta_0) n_\nu (k\eta)-j_{\nu-1} (k\eta) n_\nu (k\eta_0) \big\} \bigg].
\end{equation}
Now, let us again consider the situation $\eta \gg \eta_0$ and take the limit $\eta_0 \to 0$; $A_k (\eta)$ then becomes
\begin{equation}
A_k (\eta) = \frac{4qGU}{k^2} \left( 1-\frac{\Gamma(2\nu+1)}{2^\nu \Gamma (\nu+1)} \frac{j_{\nu-1}(k\eta)}{{(k\eta)}^{\nu-1}} \right).
\end{equation}
Correspondingly, ${\tilde F}(\eta,r)$ can be written as
\begin{equation}
{\tilde F}(\eta,r)=\frac{4qGU}{y} \int_0^\infty \frac{dz}{z^2} \left( 1-\frac{\Gamma(2\nu+1)}{2^\nu \Gamma (\nu+1)} \frac{j_{\nu-1}(z)}{z^{\nu-1}} \right) J_1 (yz), \label{tensor-sol}
\end{equation}
where $y$ is defined by $y \equiv r/\eta$.
It is clear from the last equation that ${\tilde F}$ depends on $\eta$ and $r$ only through
the ratio $r/\eta$, which means that ${\tilde F}$ exhibits self-similarity at late time
\footnote{One can arrive at ${\tilde F}$ equivalent to Eq.~(\ref{tensor-sol}) by 
making a self-similar ansatz ${\tilde F}(\eta,r)={\tilde F}(y)$ from the outset and 
by solving the resultant ordinary  differential equation for ${\tilde F}$ given by
\begin{equation}
y^2 (1-y^2) {\tilde F}''(y)+y (3+2(\nu-1) y^2) {\tilde F}'(y)+4qGU=0.
\end{equation}
Magueijo obtained the solution for the tensor-type perturbation by this approach \cite{Magueijo:1992tt}.}.
Again, for later convenience, we here give ${\tilde F}(y)$ for $\nu=1$ in terms of elementary functions;
\begin{equation}
{\tilde F}(y)=4qGU \bigg[ \frac{1}{6y^2}+\frac{1}{3} \left( \frac{3}{4} \log \frac{1+\sqrt{1-y^2}}{1-\sqrt{1-y^2}}-\sqrt{1-y^2} \left( 1+\frac{1}{2y^2} \right) \right) \Theta (1-y) \bigg]. \label{rad-full-F}
\end{equation}
%
To determine the evolution of the perturbation variables, 
Magueijo considered the closed differential equation for ${\tilde \Delta_m}$ obtained from the continuity equation ($\nabla_{\mu}T^{\mu}_{0}=0$)  \cite{Magueijo:1992tt}
\begin{equation}
{\ddot {\tilde \Delta}}_m+(1-3c_s^2) {\cal H} {\dot {\tilde \Delta}}_m-c_s^2 D^2 {\tilde \Delta}_m-\frac{3}{2} (1+2c_s^2-3c_s^4) {\cal H}^2{\tilde \Delta}_m=0. \label{master-eq-delta}
\end{equation}
whose solution in the limit $\eta \gg \eta_0$ and $r\gg c_s \eta_0$ is given by
\begin{equation}
{\tilde \Delta}_m (\eta,r) \simeq \frac{4GU (-1+(q-2) \nu)}{3\nu^2 c_s^2}{\left( 1-x^2 \right)}^{\nu-\frac{1}{2}} \Theta \left( 1-x \right). \label{sol-del}
\end{equation}
He set $q=1$ throughout and (\ref{sol-del}) coincides with the solution given in \cite{Magueijo:1992tt} in this case.
He then proceeded by solving Eqs.~(\ref{trace-part}) and (\ref{poisson-eq}) for the physically interesting case of $\nu=1$ (i.e. for a radiation-dominated Universe), to obtain 
expressions for ${\tilde \Phi}$ and ${\tilde \Psi}$.
However, the expression obtained in \cite{Magueijo:1992tt} does not satisfy the master evolution equation for ${\tilde \Psi}$, Eq.~(\ref{master-psi}), obtained from the trace part of the Einstein equations.
The inconsistency can be resolved by adding a time dependent function
which is a homogeneous solution of the Poisson Eqn. (\ref{poisson-eq}). 
The general solution of Eq.~(\ref{poisson-eq}) which is continuous and whose first derivative with respect to $r$
is continuous at $r=c_s \eta$ is 
\begin{eqnarray}
{\tilde \Phi}&=&GU \bigg[ Q(\eta)-\frac{4}{3} \log x +\frac{2}{3} \left( {\cal G}_1(x)+2\log x+\frac{8}{3}-2\log 2 \right) \Theta (1-x) \bigg], \label{gen-phi} 
\end{eqnarray}
where $Q(\eta)$ is an arbitrary function.
For the specific choice $Q(\eta)=\frac{4}{9} \left( 3\log 2-4 \right)$, this solution reduces to that given in \cite{Magueijo:1992tt}, but
the corresponding expression for ${\tilde \Psi}$, calculated from Eq.~(\ref{trace-part}), satisfies Eq.~(\ref{master-psi})
only if $Q(\eta)$ is a solution of the following differential equation;
\begin{equation}
{\ddot Q}+\frac{4}{\eta} {\dot Q}+\frac{4}{\eta^2}=0. \label{eq-Q}
\end{equation}
The general solution of Eq.~(\ref{eq-Q}) is given by
\begin{equation}
Q(\eta)=-\frac{4}{3} \left( \log \eta+\frac{4}{3}-\log 2 \right).
\end{equation}
and substituting this back into Eq.~(\ref{gen-phi}) yields an expression for ${\tilde \Phi}$ which
coincides with the one obtained previously by directly solving Eq.~(\ref{master-psi}).
\section{Properties of the perturbed spacetime}
The solutions obtained in the previous section now allow us to study the properties
of the metric perturbations excited by the string formation.
For simplicity, throughout this section, we confine our analysis to the case of a radiation-dominated universe, i.~e.~ $\nu=1$,
in which case analytic calculations are feasible to some extent.
However, we expect that the qualitative picture we obtain will hold generally true in the presence of background fluids governed by an equation of state, $P=c_s^2\rho$, with $c_s^2\neq 1/3$ and, perhaps, even in more exotic cases.
\\ \indent
The tensor-type and curvature-type metric perturbations given by Eqs.~(\ref{full-psi}) and (\ref{rad-full-F}) contain step functions
which are discontinuous at the sound horizon $r=c_s \eta$ and the cosmological horizon $r=\eta$, which implies that the properties of the perturbed spacetime differ qualitatively within each region.
We therefore split the spacetime into three regimes, I) outside the cosmological horizon $r>\eta$,
II) between the cosmological and the sound horizons $c_s \eta<r<\eta$
and III) inside the sound horizon $r<c_s \eta$, and consider the properties of the perturbed spacetime
for each regime separately.
\subsection{(I) Outside the cosmological horizon}
In this regime, the perturbed metric is given by
\begin{equation}
ds^2=\eta^2 \bigg[ -\left(1+2\Phi_I \right) d\eta^2+\left( 1+\Phi_I+F_I \right) dr^2+r^2 \left( 1+\Phi_I-F_I \right) d\theta^2+\left( 1+\Phi_I \right) dz^2 \bigg],
\end{equation}
where $\Phi_I$ and $F_I$ are defined according to
\begin{equation} \label{eq:pert_var}
\Phi_I=-\frac{4qGU}{3} \log r,~~~~~F_I=\frac{2qGU\eta^2}{3r^2}.
\end{equation}
However, performing the coordinate transformation $(\eta,r) \rightarrow (\eta(T,R),~r(T,R))$, where
\begin{equation}
\eta=T+\frac{2qGU}{3} T\log R,~~~~~r=R+\frac{qGU T^2}{3R}, \label{gauge-trans}
\end{equation}
we find that this metric reduces to the unperturbed FLRW metric,
\footnote{Since the solutions are correct only up to first order
in $GU$, we must keep in mind that we neglect any second or higher order terms in $GU$ in the new metric.}
\begin{equation}
ds^2=T^2 \left( -dT^2+dR^2+R^2 d\theta^2+dz^2 \right).
\end{equation}
It can be confirmed that the density perturbation and velocity perturbation of the perfect fluid
also vanish in the new coordinates.
Therefore, despite the initially highly noncausal appearance of the perturbation variables (\ref{eq:pert_var}), we see that information regarding the string formation does not travel beyond the light cone, and the FLRW metric remains valid beyond the causal horizon.
One may wonder how it is possible for the gauge-invariant scalar and tensor-type 
perturbations, which were nonvanishing in the original coordinates, to give vanishing perturbations beyond the light cone in the new coordinates.
In fact, it is true that, even after the coordinate transformation, neither the scalar-type perturbations, nor the tensor-type perturbations vanish individually.
However, in the new coordinates they are exactly equal in magnitude, but opposite in sign and thus cancel, yielding vanishing net perturbations.
As pointed out in \cite{Magueijo:1992tt}, the apparent violation of causality 
originates from a fact that the geometrical splitting of a perturbation into scalar and
tensor type is a nonlocal operation and manifest causality is recovered after
all types of perturbations are taken into account.
\subsection{(II) Between the sound and the cosmological horizon}
\begin{figure}[t]
  \begin{center}{
    \includegraphics[scale=0.7]{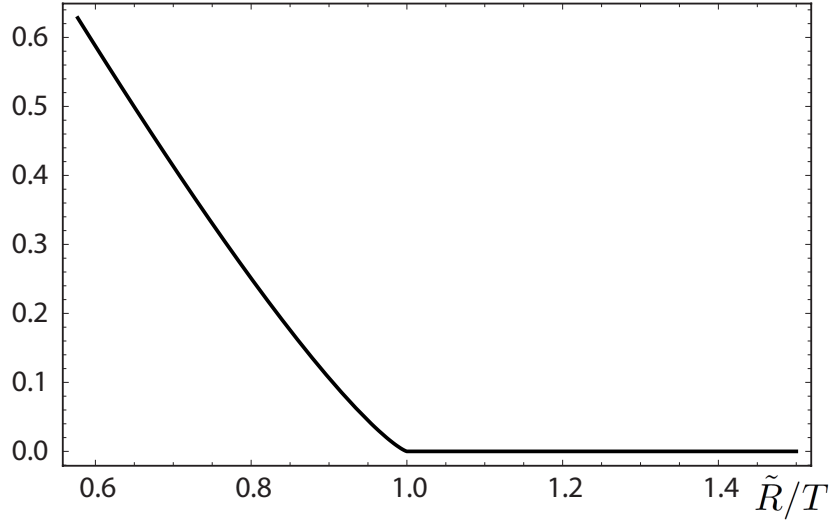}
    }
  \end{center}
  \caption{Plot of $h({\tilde R})$ outside the sound horizon using the normalization $\sqrt{qGU}T=1$.}
 \label{fig:embed}
\end{figure}

In the middle regime, the metric can be written as
\begin{equation}
ds^2=\eta^2 \bigg[ -\left(1+2\Phi_I \right) d\eta^2+\left( 1+\Phi_I+F_{II} \right) dr^2+r^2 \left( 1+\Phi_I+F_{II}+rF_{II}' \right) d\theta^2+\left( 1+\Phi_I-2F_{II}-rF_{II}' \right) dz^2 \bigg], 
\end{equation}
where $'$ indicates the derivative with respect to $r$ and $F_{II}$ is defined by
\begin{equation}
F_{II}=4qGU \bigg[ \frac{1}{6y^2}+\frac{1}{3} \left( \frac{3}{4} \log \frac{1+\sqrt{1-y^2}}{1-\sqrt{1-y^2}}-\sqrt{1-y^2} \left( 1+\frac{1}{2y^2} \right) \right) \bigg],~~~~~y=\frac{r}{\eta}.
\end{equation}
Since the matter excitation should be absent outside the sound horizon, 
we expect that only gravitational degrees of freedom, i.~e.~ tensor-type perturbations,
exist in this regime.
Given that all information regarding the tensor-type perturbations is contained within the spatial components of the metric, 
it would be convenient to perform a coordinate transformation for which the resultant, new metric,
has only spatial perturbations.
In fact, this is already achieved by the coordinate transformation given in Eq.~(\ref{gauge-trans}), which yields
\begin{equation}
ds^2=T^2 \bigg[ -dT^2+(1+{\cal G}_{II}) dR^2+R^2 (1+{\cal G}_{II}+R{\cal G}_{II}')d\theta^2+(1-2{\cal G}_{II}-R{\cal G}_{II}')dz^2 \bigg], \label{metric-middle-regime}
\end{equation}
where ${\cal G}_{II}$ is defined as
\begin{equation}
{\cal G}_{II}=-GUq \frac{2 \sqrt{1-\kappa^2} \left(2 \kappa^2+1\right)+6 \kappa^2 \log \kappa -3 \kappa^2 \log \left(-\kappa^2+2
   \sqrt{1-\kappa^2}+2\right)}{3 \kappa^2}, 
\end{equation}
and $\kappa \equiv \frac{R}{T}$ for the metric perturbations.
The density perturbation and velocity perturbation of the perfect fluid still vanish in the new coordinates.
As expected, the metric perturbations in the new coordinates still satisfy the transverse and traceless conditions,
showing explicitly that only tensor degrees of freedom are excited in this regime. 
It is interesting, however, that they depend on $T$ and $R$ only through the ratio $\kappa$, which means that the gravitational waves caused by string formation exhibit self-similarity, i.e.~waves at $(T,R)$ 
look the same as those at $(bT,bR)$, where $b$ is a numerical constant \cite{Harada}.
Magueijo assumed a self-similar ansatz in this regime and our results confirm that this is correct.
However, \cite{Magueijo:1992tt} also states that the gravitational potential in the region $c_s \eta \le r \le \eta$
is that of a static repulsive rod, whereas the metric (\ref{metric-middle-regime}) shows that a potential 
is not present between the sound horizon and the cosmological horizon.
Now let us consider the embedding of the two-dimensional surface spanned by the $R$ and $\theta$
coordinates at fixed $T$ and $z$.
The line element of the subspace is given by
\begin{equation}
ds_{(2)}^2=T^2 \bigg[(1+{\cal G}_{II}) dR^2+R^2 (1+{\cal G}_{II}+R{\cal G}_{II}')d\theta^2 \bigg].
\end{equation}
Introducing a new radial coordinate ${\tilde R}$,
\begin{equation}
{\tilde R}=R \left( 1+\frac{1}{2}{\cal G}_{II}+\frac{R}{2} {\cal G}_{II}' \right),
\end{equation}
we find that this can be rewritten as
\begin{equation}
ds_{(2)}^2=T^2 \bigg[(1+4qGU \sqrt{1-{\tilde \kappa}^2}) d{\tilde R}^2+{\tilde R}^2d\theta^2 \bigg],~~~~~{\tilde \kappa}=\frac{\tilde R}{T}.
\end{equation}
Therefore, $2\pi {\tilde R}$ is the circumference of a comoving circle whose radial coordinate is
specified by ${\tilde R}$.
The height of the embedding surface in the three-dimensional space, relative to that for the region outside the light cone, which we denote $h({\tilde R})$, is then given by
\begin{equation}
h({\tilde R})=2\sqrt{qGU}T \bigg[ \frac{\sqrt{\pi} \Gamma( \frac{5}{4} )}{2 \Gamma (\frac{7}{4})}-\frac{\tilde R}{T} F \left( -\frac{1}{4},\frac{1}{2},\frac{3}{2};\frac{{\tilde R}^2}{T^2} \right) \bigg].
\end{equation}
We find that the first derivative of $h({\tilde R})$ at $T={\tilde R}$ vanishes but
the second derivative diverges. 
A plot of $h({\tilde R})$ is given in Fig.~\ref{fig:embed}.
Next, let us consider how the infinitesimal distance between two particles following initially parallel geodesics
changes as they across the gravitational wave front at the cosmological horizon.
The change of distance is described by the geodesic deviation equation (see, for example, Wald \cite{wald}),
\begin{equation}
\frac{D^2 \xi^\mu}{d\tau^2}=R^\mu_{~\alpha \beta \nu} \frac{dx^\alpha}{d\tau} \frac{dx^\beta}{d\tau} \xi^\nu,
\end{equation}
where $\xi^\mu$ is the infinitesimal displacement vector between the two particles and $\tau$ is the proper time.\footnote{
The Riemann tensor is defined by $R^\rho_{~\sigma \mu \nu}=\partial_\mu \Gamma^\rho_{~\nu \sigma}-\partial_\nu \Gamma^\rho_{~\mu \sigma}+\Gamma^\rho_{~\mu \lambda} \Gamma^\lambda_{~\nu \sigma}-\Gamma^\rho_{~\nu \lambda} \Gamma^\lambda_{~\mu \sigma}$.}
Assuming that the particles are initially attached to the comoving coordinates when
they are outside the light cone, the component $R^i_{~0 0 i}$ gives the dominant force for the
displacement along $i$-th axis. The relevant components are therefore,
\begin{eqnarray}
R^R_{~00R}&=&-\frac{1}{T^2}-\frac{2q GU \left(\kappa^2+2 \right)\sqrt{1-\kappa^2}}{3 T^2 \kappa^2}, \\
R^\theta_{~00\theta}&=&-\frac{1}{T^2}+\frac{2q GU \left(2 \kappa^4-\kappa^2+2\right)}{3 T^2 \kappa^2 \sqrt{1-\kappa^2}}, \\
R^Z_{~00Z}&=&-\frac{1}{T^2}-\frac{2q GU \kappa^2}{T^2 \sqrt{1-\kappa^2}}.
\end{eqnarray}
The first term, $-1/T^2$, common to all the components above is merely the tidal force
due to the cosmological expansion. It is simply the second time derivative of the scale factor and the minus sign is due to decelerating expansion of the Universe. 
The second term in each component represents the additional tidal force induced by
the gravitational waves.
We find that, at the cosmological horizon where $\kappa=1$, the $R$-component of the Riemann tensor vanishes
while the $\theta$ and $Z$-components diverge toward positive and negative infinity, respectively.
There is therefore a gravitational shock front at the edge of the light cone and, although the tidal force diverges, this does not imply that two particles
separated by a small distance undergo a large displacement as they cross the
cosmological horizon.
The divergence is weak enough to make the solution of the geodesic deviation equation,
i.e. the double time-integration of the tidal force, finite.
\begin{figure}[t]
  \begin{center}{
    \includegraphics[scale=0.7]{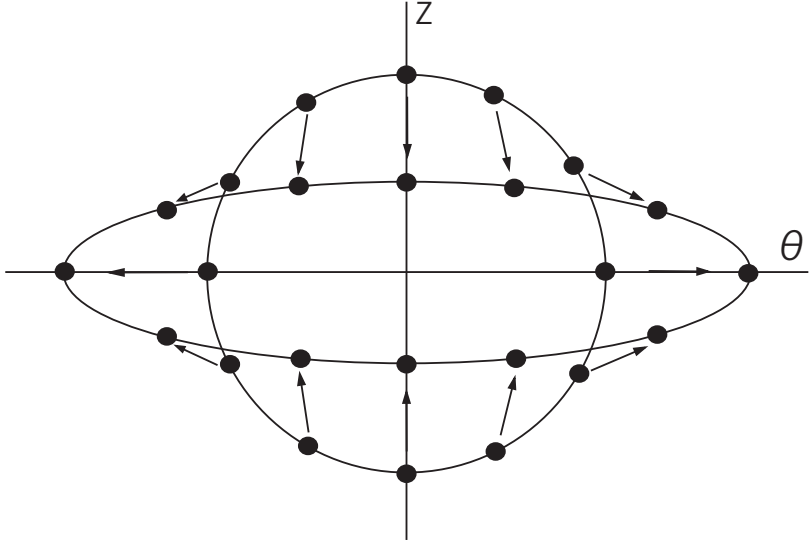}
    }
  \end{center}
  \caption{The motion of free particles after crossing the cosmological horizon.}
 \label{fig:motion}
\end{figure}
The motion of a ring of free particles hit by the gravitational shock wave at the cosmological horizon is 
schematically plotted in Fig.~\ref{fig:motion}.
The reason that the ring is elongated along the angular direction as the particles cross the cosmological
horizon is intuitively understood as follows.
Let us consider a ring which consists of free particles with the same radial coordinate $R_0$, separated by equal angles,
and vary $R_0$ from $R_0 >T$ to $R_0 <T$.
This procedure represents how the ring is distorted as it crosses the cosmological horizon.
Before the ring hits the cosmological horizon, the ratio of the rate of decease of the circumference of the ring to that of the radius is given by $2\pi$
and the particles move closer together at a constant rate.
After the ring crosses the horizon, due to the distortion of space caused by the gravitational
wave, the rate of decrease of the physical radius (which is the length of the solid line in 
Fig.~\ref{fig:embed}), becomes greater than that of the circumference.
This can be understood by considering the limiting case where the slope of the curve for ${\tilde R}/T$
in Fig.~\ref{fig:embed} is very steep.
In this case, even a significant translation along the solid line results in a tiny shift in ${\tilde R}$.
Therefore, the rate at which particles on the ring move closer to each other is slower inside the horizon, than it is outside, 
which is equivalent to saying that the particles have been affected by the outward tidal force that 
stretches the distance between two particles.
\subsection{(III) Inside the sound horizon}
In the final regime we consider the region of spacetime inside the sound cone, in which the metric can be written as
\begin{equation}
ds^2=\eta^2 \bigg[ -\left(1+2\Phi_{III} \right) d\eta^2+\left( 1+2 \Psi_{III} \right) \big\{ \left( 1+F_{II} \right) dr^2+r^2 \left( 1+F_{II}+rF_{II}' \right) d\theta^2+\left( 1-2F_{II}-rF_{II}' \right) dz^2 \big\} \bigg],
\end{equation}
where $\Psi_{III}$ is defined as
\begin{equation}
\Psi_{III}=GU \bigg[ -2 \log r+\frac{(q-3)}{3} (-2 \log \eta+{\cal G}_1 (x) ) \bigg],
\end{equation}
and $\Phi_{III}$ is then given by Eq.~(\ref{trace-part}).
We find that the following coordinate transformation,
\begin{eqnarray}
\eta&=&T-\frac{3GU \left(R^2-(q-1) Z^2\right)}{2 T}-\frac{GU}{6} (q-3) T (2 \log T-1), \\
r&=&R+GU(q+2) R \log R-\frac{GU}{6} R (2q (2+3 \log 2)+18 \log T+9), \\
z&=&Z+\frac{GU}{6} Z \left(3-13q+6q \log 4+ 18(q-1) \log T \right),
\end{eqnarray}
yields a new metric which reduces to 
\begin{equation}
ds^2 \approx T^2 \bigg[ \big\{ -1+4GU (q-1) \log R \big\} dT^2+dR^2+R^2 \big\{ 1-4GU (1+q) \big\} d\theta^2+\big\{ 1+4GU (q-1) \log R \big\} dZ^2 \bigg],
\end{equation}
in the subhorizon limit, $R/T \ll 1$. 
This is the metric for an eternal cosmic string and coincides with the one given by Vilenkin, Gott and Hiscock in previous studies \cite{167218, Gott1, Hiscock1}.
Therefore, as expected, the spacetime sufficiently close to a noneternal cosmic string resembles that for the eternal string. 
The nonvanishing time-time component of the metric perturbation leads to a gravitational acceleration 
for test particles given by $-\frac{2GU(1-q)}{R}$.
It is therefore an attractive force for $q<1$ and becomes repulsive for $q>1$,
which also makes sense intuitively since $0 \le q\ll 1$ corresponds to a nonrelativistic string and $q>1$
corresponds to a tension-dominated string.
For the Nambu-Goto string, for which $q=1$, there is no gravitational force.
In all cases, there is a deficit angle around the string of $4\pi GU (1+q)$.   
\\ \indent
The general form of the metric in the new coordinate system, without the subhorizon approximation, is very complicated.
For simplicity, we here consider only the Nambu-Goto string, for which $q=1$.
In this case, the metric perturbations, which are valid in the whole region inside the sound horizon,
can be written as
\begin{eqnarray}
h_{TT,(III)}&=&\frac{8}{9} GU \left(\sqrt{1-x^2} x^2-4 \sqrt{1-x^2}+3 \log \left(\sqrt{1-x^2}+1\right)+4-\log 8 \right), \\
h_{RR,(III)}&=&\frac{GU}{9 x^2} \bigg[ -6 \sqrt{9-3 x^2}-x^2 \left(4 \sqrt{9-3 x^2}+32 \sqrt{1-x^2}-41+27 \log 3+42 \log 2 \right) \nonumber \\
&&+3 x^2 \left(3 \log \left(-x^2+2 \sqrt{9-3 x^2}+6\right)+8 \log
   \left(\sqrt{1-x^2}+1\right)+6\log 3 \right)+\left(8 \sqrt{1-x^2}-9\right) x^4+18 \bigg], \\
h_{\theta \theta,(III)}&=&\frac{GU}{9x^2} \bigg[ 6 \left(\sqrt{9-3 x^2}-3\right)-x^2 \left(8 \sqrt{9-3 x^2}+32 \sqrt{1-x^2}+13+9 \log 3+42 \log 2\right) \nonumber \\
&&+3 x^2 \left(3 \log \left(-x^2+2 \sqrt{9-3 x^2}+6\right)+8 \log
   \left(\sqrt{1-x^2}+1\right)\right)+\left(8 \sqrt{1-x^2}-9\right) x^4\bigg], \\
h_{ZZ,(III)}&=&\frac{1}{9} GU \bigg[ \left(8 \sqrt{1-x^2}-9\right) x^2+2 \left(6 \sqrt{9-3 x^2}-16 \sqrt{1-x^2}-2+6\log 2+9\log 3\right) \nonumber \\
&&-18 \log \left(-x^2+2 \sqrt{9-3 x^2}+6\right)+24 \log
   \left(\sqrt{1-x^2}+1\right)\bigg],     
\end{eqnarray}
and all other components vanish.
It is interesting to see that all the components of the metric perturbation depend on $T$ and $R$ only through $x$.
Therefore, as in the case for the middle regime, the perturbations also exhibit self-similarity
inside the sound cone.
Unlike when we approach the edge of the light cone, we find that the Riemann tensor does not diverge at the sound horizon and the relevant components are given by
\begin{equation}
R^R_{~00R}=-\frac{1}{T^2}-\frac{2(15+7\sqrt{6})GU}{9T^2},~~~R^\theta_{~00\theta}=-\frac{1}{T^2}+\frac{(17\sqrt{6}-30)GU}{9T^2},~~~R^Z_{~00Z}=-\frac{1}{T^2}-\frac{(10+\sqrt{6})GU}{3T^2}.
\end{equation}
Sufficiently close to the string core, these reduce to
\begin{equation}
R^R_{~00R}=-\frac{1}{T^2}+\frac{6GU}{T^2}+{\cal O}(x^2),~~~R^R_{~00R}=-\frac{1}{T^2}+\frac{6GU}{T^2}+{\cal O}(x^2),~~~R^Z_{~00Z}=-\frac{1}{T^2}+{\cal O}(x^2).
\end{equation}
There therefore remains a residual outward tidal force along both the radial and angular directions,
which does not vanish even in the limit $x\to 0$ but which, of course, still vanishes in the eternal limit, $T \to \infty$.
\section{Conclusion}
We derived the analytic form of the metric perturbations on an FLRW background, excited
by the formation of a straight cosmic string under the wire approximation.
Our results are valid for a general string with differing tension and energy per unit length.
For the Nambu-Goto string, our metric does not coincide with that given in previous work due to 
a differing choice of boundary conditions for the Poisson equation.
We found that, as a result of this choice, the solutions given in \cite{Magueijo:1992tt}  for the gravitational potential and curvature perturbation variables do not satisfy the trace part of the Einstein equations.
\\ \indent
By performing the appropriate coordinate transformation, we explicitly verified that the spacetime outside
the light cone remains unperturbed and governed by the FLRW metric.
Whereas it is nontrivial to demonstrate nonviolation of causality in the original metric, 
it is manifestly respected in the new coordinate system.
At the cosmological horizon, we found that there exists a gravitational shock giving rise to a divergent tidal force
along both the angular direction and the direction of the string axis, $z$.
The shock distorts a small ring of test particles placed in the $\theta-z$ plane by stretching it along $\theta$-direction
and shrinking it along the $z$-direction.
The degree of divergence is weak, so that the resultant distortion remains finite.
Inside the cosmological horizon, the metric perturbations exhibit self-similarity at late time,
i.e. they depend only on a ratio of the radial coordinate to the conformal time.
This is reasonable since the Hubble distance is the only characteristic length scale in the system.
Deep inside the horizon, we found that the metric tends to that for an eternal string at late times,
which has been used to search for the signatures of cosmic strings in astronomical observations. 
\begin{acknowledgments}
This work was supported by JSPS Research Fellowships and Grants-in-Aid for JSPS Fellows No.~1008477 (TS) and No.~P10327 (ML).
\end{acknowledgments}
\appendix
%
%
\section{Formula for integral of a product of two Bessel functions}
In this appendix, we provide a formula for the integral $I$ defined by 
\begin{equation}
I=\int_0^\infty x^\lambda J_\mu (ax) J_\nu (bx).
\end{equation}
For $a>b>0$, we have
\begin{equation}
I=\frac{2^\lambda b^\nu \Gamma (\alpha)}{a^{\lambda+\nu+1} \Gamma (\nu+1) \Gamma (1-\beta) } F\left( \alpha,\beta,\nu+1;\frac{b^2}{a^2} \right),
\end{equation}
where $F$ is the usual Gauss hypergeometric function and $\alpha$ and $\beta$ are defined by
\begin{equation}
\alpha=\frac{1}{2} (\lambda+\mu+\nu+1),~~~\beta=\frac{1}{2} (\lambda-\mu+\nu+1).
\end{equation}
For $a=b>0$, we have
\begin{equation}
I=\frac{2^\lambda \Gamma (\alpha) \Gamma (-\lambda)}{a^{\lambda+1} \Gamma (1-\beta) \Gamma (\nu-\alpha+1) \Gamma (\nu-\beta+1)}.
\end{equation}
and for $b>a>0$, 
\begin{equation}
I=\frac{2^\lambda a^\mu \Gamma (\alpha)}{b^{\lambda+\mu+1} \Gamma (\mu+1) \Gamma (\nu-\alpha+1)} F \left( \alpha,\alpha-\nu,\mu+1;\frac{a^2}{b^2} \right).
\end{equation}


\begin{thebibliography}{100}
%
\bibitem{Nielsen_Olesen}
  H.~B.~Nielsen and P.~Olesen,
  Nucl.\ Phys.\  {\bf B61}, 45-61 (1973). 
  %
\bibitem{VS1994} 
  A.~Vilenkin and E.~P.~S.~Shellard,
   {\it Cosmic strings and other topological defects, in Cambridge Monographs in Mathematical Physics} (Cambridge University Press, Cambridge, England, 2000). 
  %
\bibitem{hep-ph/9411342} 
  M.~B.~Hindmarsh and T.~W.~B.~Kibble,
  Rept.\ Prog.\ Phys.\ \ {\bf 58}, 477  (1995),
  arXiv:hep-ph/9411342.
   %
\bibitem{hep-th/0508135v2}
  A.~Vilenkin, 
  Frontiers in Science Series (2006),
  arXiv:arXiv:hep-th/0508135v2.
  %
\bibitem{Preskill}
  J.~Preskill
  Lectures presneted at the 1985 Les Houches Summer School
  US D.O.E. Research and Development Report, CALT-66-1287 (1986).
  %
 \bibitem{Goto1}
  T.~Goto
  Prog. Theor. Phys. {\bf 46}, 1560 (1971).
  %
\bibitem{Anderson2003} 
  M.~R.~Anderson,
  {\it The Mathematical Theory of Cosmic Strings: Cosmic Strings in the Wire Approximation} (Institute of Physics, London, 2003).
   %
 \bibitem{Nambu1}
  Y.~Nambu
  Nucl. Phys. {\bf B130}, 505 (1977).
  %
\bibitem{hep-ph/0308134} 
  R.~Jeannerot, J.~Rocher and M.~Sakellariadou,
  Phys.\ Rev.\ D\ {\bf 68}, 103514  (2003),
  [hep-ph/0308134].
  %
 \bibitem{Abrikosov1}
 A~.A.~Abrikosov,
 Sov.\ Phys.\ JETP {\it 5}, 1174 (1957).
  %
 \bibitem{Zurek1}
 W.H.~Zurek,
 Nature (London) {\bf 317}, 505-508 (1985).
%
 \bibitem{Zurek2}
 W.H.~Zurek,
 Phys. Rept. {\bf 276}, 177-221 (1996),
 arXiv:cond-mat/9607135v1.
%
 \bibitem{Zurek3}
 W.H.~Zurek,
 arXiv:cond-mat/9502119 (1995).
%
 \bibitem{Bowick1}
 M.J.~Bowick, L.~Chandar, E.A.~Schiff, A.M.~Srivastava,
 Science {\bf 263}, 943-945 (1994),
 arXiv:hep-ph/9208233v1.
%
 \bibitem{Williams1}
 G.A.~Williams,
 Phys. Rev. Lett. {\bf 82}, 1201 (1999). 
 arXiv:cond-mat/9807338v3
%
 \bibitem{Hendry1}
 P.C.~Hendry, N.S.~Lawson, R.A.M.~Lee, P.V.E.~McClintock and C.D.H.~Williams,
 J. Low Temp. Phys. {\bf 93}, 5-6, 1059-1067 (1993).
%
 \bibitem{Chuang1}
 I.~Chuang, R.~Durrer, N.~Turok, B.~Yurke,
 Science {\bf 251}, 1336-1342 (1991).  
 %
 \bibitem{Annett1}
 J.F.~Annett, 
 {\it Superconductivity, Superfluids and Condensates} Oxford University Press, New York, 2004). 
   %
\bibitem{ICTP/75/5} 
  T.~W.~B.~Kibble,
  J.\ Phys.\ AA\ {\bf 9}, 1387  (1976).
 %
\bibitem{Witten:1985}
  E.~Witten,
  Phys.\ Lett. {\bf B153}, 243 (1985).
  %
 \bibitem{Polchinski_Intro}
  J.~Polchinski,
 arXiv:hep-th/0412244.
 %
\bibitem{0911.1345v3} 
  E.~J.~Copeland and T.~W.~B.~Kibble,
 Proc.\ Roy.\ Soc.\ Lond. {\bf A466}:623-657 (2010)
  arXiv:0911.1345v3 [hep-th]
\bibitem{CMP_FD1}
  E.~J.~Copeland, R.~C.~Myers and J.~Polchinski,
  J. High Energy Phys. {\bf 06}, 013 (2004),
  arXiv:hep-th/0312067.
   %
\bibitem{CMP_FD2}
  E.~J.~Copeland, R.~C.~Myers and J.~Polchinski,
  Comptes Rendus Physique {\bf 5}, 9-10, 1021-1029 (2004).
  %
\bibitem{0811.1277v1}  
  A.~Achucarro and C.~J.~A.~P.~Martins,
  {\it Encyclopedia of Complexity and Systems Science} (Springer,   
  New York, 2009),
  arXiv:0811.1277v1 [astro-ph].
  %
\bibitem{hep-th/0505050v1} 
  A.~Davis and T.~W.~B.~Kibble,
  Contemp.\ Phys. {\bf 46}, 313-322 (2005),
  arXiv:hep-th/0505050v1.
  %
\bibitem{astro-ph/0410073v2} 
  T.~W.~B.~Kibble,
  arXiv:astro-ph/0410073v2 (2004).
  %
\bibitem{Sakellariadou:2009}
  M.~Sakellariadou,
 Nucl.\ Phys.\ B,\ Proc.\ Suppl. {\bf 192}, 68 (2009),
  arXiv:0902.0569 [hep-th].
 %
\bibitem{Rajantie:2007}
  A.~Rajantie, M.~Sakellariadou and H.~Stoica,
  J. Cosmol. Astropart. Phys. {\bf 0711}, 021 (2007),
  arXiv:0706.3662 [hep-th].
  %
()\cite{Copeland:2005}
\bibitem{Copeland:2005}
 E.~J.~Copeland and P.~M.~Saffin,
 J. High Energy Phys. {\bf 0511}, 023 (2005),
arXiv:hep-th/0505110.
  %
\bibitem{Cline1}
   J.~M.~Cline, 
  arXiv:hep-th/0612129 (2006).
  %
\bibitem{Gasperini1}
 M.~Gasperini, 
 {\it Elements of String Cosmology} (Cambridge University Press, Cambridge, England, 2006).
 %
\bibitem{Tye1}
 S.-H.H.~Tye
hep-th/0610221v2 (2006). 
 %
 \bibitem{Carroll:TASI}
  S.M.~Carroll
 arXiv:hep-th/0011110v2 (2000).
 %
 \bibitem{Quevedo:Lectures}
  F.~Quevedo
  Class. Quant. Grav. {\bf 19}, 5721-5779 (2002),
 arXiv:hep-th/0210292v1.
 %
 \bibitem{Danielsson1}
  U.H.~Danielsson
  Class. Quant. Grav. {\bf 22}, S1-S40 (2005),
 arXiv:hep-th/0409274.
 %
 \bibitem{Sarangi1}
  S.~Sarangi and S.-H.~Tye,
 Phys. Lett. B {\bf 536}, 185 (2002),
 arXiv:hep-th/0204074v1.
 %
 \bibitem{Jones_etal1}
  N.T.~Jones, H.~Stoica and S.-H.~Tye,
 Phys. Lett. {\bf B563}, 6 (2003),
 arXiv:hep-th/0303269v1].
 %
 \bibitem{Pogosian_Obs1}
  L.~Pogosian, S.~-H.~Tye, I.~Wasserman and M.~Wyman, 
 Phys.\ Rev.\ D {\bf 68}, 023506 (2003),
 arXiv:hep-th/0304188.
 %
 \bibitem{Avgoustidis:2004}
  A.~Avgoustidis and E.~P.~S.~Shellard,
  Phys.\ Rev.\ D {\bf 71}, 123513 (2005),
  arXiv:hep-ph/0410349.
    %
  %
\bibitem{Leblond:2007}
  L.~Leblond and M.~Wyman,
  Phys.\ Rev.\ D {\bf 75}, 123522 (2007),
  arXiv:astro-ph/0701427.
    %
\bibitem{Dasgupta1}
  K.~Dasgupta, H.~F.~Firouzjahi and R.~Gwyn,
  J. High Energy Phys.  {\bf 0704}, 093 (2007),
  arXiv:hep-th/0702193v3.
  %
\bibitem{Martins1}
  C.~J.~A.~P.~Martins,
  Phys.\ Rev.\   D {\bf 82}, 067301 (2010),
  arXiv:1009.1707v1 [hep-ph].  
%
%
\bibitem{Matsuda2}
  T.~Matsuda,
  J. High Energy Phys. {\bf 0505}, 015 (2005),
  arXiv:hep-ph/0412290.
%
%
%
\bibitem{Lake1}
M.~Lake, S.~Thomas and J.~Ward,
 J. High Energy Phys. {\bf 12} (2009) 033,
 arXiv:0906.3695v2 [hep-ph].
%
%
\bibitem{Lake3}
M.~Lake and J.~Ward,
 J. High Energy Phys. {\bf 04} (2011) 048,
arXiv:1009.2104 [hep-ph].
%
\bibitem{Siemens1}
  X.~Siemens, X.~Martin and K.~D.~Olum,
  Nucl.\ Phys.\  {\bf B595}, 402-414 (2001),
  arXiv:astro-ph/0005411v1.
  %
\bibitem{Blanco-Pillado1}
  J.~J.~Blanco-Pillado and K.~D.~Olum,
  J. Cosmol. Astropart. Phys.  {\bf 1005}, 014 (2010),
  arXiv:0707.3460v3 [astro-ph].
  %
\bibitem{vilenkin} 
  A.~Vilenkin,
  Phys.\ Rev.\ D\ {\bf 41}, 3038  (1990).
  %
\bibitem{carter} 
  B.~Carter,
  Phys.\ Rev.\ D\ {\bf 41}, 3869  (1990).
  %
  %
\bibitem{witten} 
  E.~Witten,
  Nucl.\ Phys.\ B\ {\bf 249}, 557  (1985).
    %
\bibitem{FPRINT-92-39} 
  P.~Peter,
  Phys.\ Rev.\ D\ {\bf 45}, 1091  (1992).
%
\bibitem{167218} 
  A.~Vilenkin,
  Phys.\ Rev.\ D\ {\bf 23}, 852  (1981).
    %
\bibitem{Gott1}
  J.~R.~Gott III,
  Astrophysical.\ J.\  {\bf 288}: 422-427 (1985).
  %
\bibitem{Hiscock1}
  W.~A.~Hiscock, 
 Phys.\ Rev.\ D {\bf 31}, 3288-3290 (1985).
 %
\bibitem{Pe94}
  P.~Peter, 
 Class.\ Quantum.\ Grav. {\bf 11}, 131-137 (1994).
   %
\bibitem{Uz01}
  J-P.~Uzan and F.~Bernardeau, 
 Phys.\ Rev.\ D {\bf 63}, 023004 (2001).
%
%
%
%
\bibitem{Vilenkin_lensing1}
  A.~Vilenkin,
  Astrophysical.\ J.\  {\bf 282}:L51-L54 (1984).
  %
\bibitem{Hogan_lensing1}
  C.~Hogan and R.~Narayan,
  Mon.\ Not.\ R.\ astr.\ Soc.  {\bf 211}, 575-591 (1984).
  %
\bibitem{Shlaer_lensing1}
  B.~Shlaer and M.~Wyman,
  Phys.\ Rev.\  D {\bf 72}, 123504 (2005),
  arXiv:hep-th/0509177v1.
  %
\bibitem{Huterer_lensing1}
  D.~Huterer and T.~Vachaspati,
  Phys.\ Rev.\  D {\bf 68}, 041301 (2003), 
  arXiv:astro-ph/0305006v1.
    %
     %
\bibitem{Hindmarsh_lensing1}
  M.~Hindmarsh and A.~Wray,
  Phys.\ Lett.\  {\bf B251}, 4, 498-502 (1990).  
    %
\bibitem{Dyda_lensing1}
  S.~Dyda and R.~H.~Brandenberger,
  arXiv:0710.1903v1 [astro-ph].
    %
\bibitem{Gasperini_lensing1}
  M.~A.~Gasperini and P.~Marshall,
  Mon.\ Not.\ R.\ astr.\ Soc.  {\bf 385}, 4, 1959-1964 (2008).
  %
\bibitem{Morganson_lensing1}
  M.~Morganson, P.~Marshall, T.~Treu, T.~Schrabback and R.~D.~Blandford,
  Mon.\ Not.\ R.\ astr.\ Soc.  {\bf 406}, 4, 2452-2472 (2010).
    %
\bibitem{deLaix_lensing1}
  A.~A.~de Laix, L.~M.~Kraus and T.~Vachaspati,
  Phys.\ Rev.\ Lett.  {\bf 79}, 1968-1971 (1997),
  arXiv:astro-ph/9702033v1.
  %
%
%
\bibitem{Cowie_lensing1}
  L~L.~Cowie and E.~M.~Hu,
 Astrophysical.\ J. {\bf 318}:L33-L38 (1987).
 %
\bibitem{deLaix_lensing2}
  A.~A.~de Laix and T.~Vachaspati,
 Phys.\ Rev. D {\bf 54}, 4780-4791 (1996),
  arXiv:astro-ph/9605171.
  %
\bibitem{Mack_lensing1}
  K.~J.~Mack, D.~H.~Wesley and L.~J.~King,
 Phys.\ Rev. D {\bf 76}, 123515 (2007),
  arXiv:astro-ph/0702648.
%
%
%
\bibitem{Zeldovich_CMB1}
  Ya.~B.~Zeldovich, 
 Mon.\ Not.\ R.\ astr.\ Soc.  {\bf 192}, 663-667 (1980).
 %
\bibitem{Kaiser_CMB1}
  N.~Kaiser and A.~Stebbins, 
 Nature (London) {\bf 310}, 391-393 (1984).
 %
\bibitem{Stebbins_CMB1}
  A.~Stebbins,
  Astrophys.\ J. {\bf 327}, 584 (1988).
     %
\bibitem{Gangui_CMB1}
  A.~Gangui and L.~Perivolaropoulos,
  Astrophys.\ J. {\bf 447}: 1-7 (1995),
  arXiv:astro-ph/9408034v1.
  %
\bibitem{Allen_CMB1}
  B.~Allen, R.~R.~Caldwell, E.~P.~S.~Shellard, A.~Stebbins and V.~Veeraraghavan,
   Phys.\ Rev.\ Lett. {\bf 77}, 3061-3065 (1996),
   arXiv:astro-ph/9609038v1.
      %
\bibitem{Perivolaropoulos_CMB1}
  L.~Perivolaropoulos,
  arXiv:astro-ph/9704011v1 (1997).
      %
\bibitem{Benabed_CMB1}
  K.~Benabed and F.~Bernardeau,
  Phys.\ Rev. D {\bf 61}, 123510 (2000),
  arXiv:astro-ph/9906161v2.
 %
\bibitem{Landriau_CMB1}
  M.~Landriau and E.~P.~S.~Shellard, 
 Phys.\ Rev.\ D {\bf 69}, 023003 (2004),
 arXiv:astro-ph/0302166.
  %
\bibitem{Landriau_CMB2}
  M.~Landriau and E.~P.~S.~Shellard, 
 Phys.\ Rev.\ D {\bf 83}, 043516 (2011),
 arXiv:1004.2885 [astro-ph.CO].
  %
\bibitem{Pogosian_CMB1}
  L.~Pogosian, S.~-H.~Tye, I.~Wasserman and M.~Wyman, 
 J. Cosmol. Astropart. Phys. {\bf 02}, 013 (2009),
 arXiv:0804.0810v2 [astro-ph].
   %
\bibitem{Wyman_CMB1}
  M.~Wyman and L.~Pogosian, 
 Phys.\ Rev. D {\bf 72}, 023513 (2005),
 arXiv:astro-ph/0503364.
   %
\bibitem{Battye_CMB1}
  R.~A.~Battye, B.~Garbrecht, A.~Moss and H.~Stoica, 
   J. Cosmol. Astropart. Phys. {\bf 01}, 020 (2008),
 arXiv:0710.1541v3 [astro-ph].
  %
\bibitem{Battye_CMB2}
  R.~A.~Battye and A.~Moss, 
 Phys.\ Rev.\ D {\bf 82}, 023521 (2010),
 arXiv:1005.0479v2 [astro-ph.CO].
   %
\bibitem{Contaldi_CMB1}
  C.~Contaldi, M.~Hindmarsh and J.~Mageuijo, 
Phys.\ Rev.\ Lett. {\bf 82}, 2034-2037 (1999),
 arXiv:astro-ph/9809053.
 %
\bibitem{Bevis_CMB1}
   N.~Bevis, M.~Hindmarsh, M.~Kunz and  J.~Urrestilla,
   Phys.\ Rev.\ D {\bf 75}, 065015 (2007),
  arXiv:astro-ph/0605018.
  %
\bibitem{Bevis_CMB2}
   N.~Bevis, M.~Hindmarsh, M.~Kunz and  J.~Urrestilla,
   Phys. Rev. D {\bf 82}, 065004 (2010),
  arXiv:1005.2663v1 [astro-ph.CO]. 
  %
\bibitem{Bevis_CMB3}
   J.~Urrestilla, P.~Mukherjee, A.R.~Liddle, N.~Bevis, M.~Hindmarsh and M.~Kunz,
   Phys.\ Rev.\ D {\bf 77}, 123005 (2008),
  arXiv:0803.2059v2 [astro-ph].
    %
\bibitem{Bevis_CMB4}
  N.~Bevis, M.~Hindmarsh, M.~Kunz and J.~Urretilla, 
 Phys.\ Rev.\ Lett. {\bf 100}, 021301 (2008),
 arXiv:astro-ph/0702223.
          %
\bibitem{Suyama_CMB1}
  M~Hindmarsh, C.~Ringeval and T.~Suyama,
  Phys.\ Rev. D {\bf 80}, 083501 (2009),
  arXiv:0908.0432 [astro-ph.CO].
         %
\bibitem{Suyama_CMB2}
  M~Hindmarsh, C.~Ringeval and T.~Suyama,
  Phys.\ Rev. D {\bf 81}, 063505 (2010),
  arXiv:0911.1241 [astro-ph.CO].
  %
  %
      %
      %
\bibitem{Bhattacharjee_CosRay1}
  P.~Bhattacharjee and N.~C.~Nana, 
  Phys.\ Lett. {\bf B246}, 3-4, 365-370 (1990).
      %
\bibitem{Sigl_CosRay1}
  G.~Sigl, 
  Space Science Reviews {\bf 75}, 1-2, 375-385 (1996),
  arXiv:astro-ph/9503014.
      %
\bibitem{Berezinsky_CosRay1}
  V.~Berezinsky and A.~Vilenkin, 
  Phys.\ Rev.\ Lett. {\bf 79}, 5202-5205 (1997),
  arXiv:astro-ph/9704257.
        %
\bibitem{Wichoski_CosRay1}
U.~F.~Wichoski, R.~H.~Brandenberger and J.~H.~MacGibbon, 
  arXiv:hep-ph/9903545v1 (1999).
  %
  %
\bibitem{Khatri_21cm}
  R.~Khatri and B.~D.~Wandelt, 
 Phys.\ Rev.\ Lett. {\bf 100}, 091302 (2008),
 arXiv:0801.4406v1 [astro-ph].
 %
  %
\bibitem{Gr09}
  J.~Greiner et al, 
  Astrophys.\ J. {\bf 693}, 1610 (2009),
  arXiv:0810.2314 [astro-ph].
 %
\bibitem{Ta09}
  N.~R.~Tanvir et al., 
  Nature (London) {\bf 461}, 1254 (2009),
  arXiv:0906.1577 [astro-ph.CO].
   %
\bibitem{Sa09}
  R.~Salvaterra et al., 
  Nature (London) {\bf 461}, 1258 (2009),
  arXiv:0906.1578 [astro-ph.CO].
    %
\bibitem{Ba87}
  A.~Babul, B.~Paczynski and D.~Spergel, 
  Astrophys.\ J. {\bf 316}, L49 (1987).
    %
\bibitem{Pa88}
  B.~Paczynski, 
  Astrophys.\ J. {\bf 335}, 525 (1988).
      %
\bibitem{Br93}
  R.~H.~Brandenberger, A.~T.~Sornborger and M.~Trodden, 
  Phys.\ Rev. D {\bf 48}, 940 (1993).
      %
\bibitem{Pl94}
  R.~Plaga, 
  Astrophys.\ J. {\bf 424}, L9 (1994).
      %
\bibitem{Be01}
  V.~Berezinsky, B.~Hnatyk and A.~Vilenkin, 
  Phys.\ Rev. D {\bf 64}, 043004 (2001),
  arXiv:astro-ph/0102366.
    %
\bibitem{Be04}
  V.~Berezinsky, B.~Hnatyk and A.~Vilenkin, 
  Baltic Astronomy, {\bf 13}, 289 (2004),
  arXiv:astro-ph/0102366.
    %
\bibitem{Ch10}
  K.~S.~Cheng, Y.-W.~Yu and T.~Harko, 
  Phys.\ Rev.\ Lett. {\bf 104}, 241102 (2010),
  arXiv:1005.3427v2 [astro-ph.HE].
   %
\bibitem{Wa11}
  Y.~Wang, Y.-Z.~Fan and D.-M.~Wei, 
  Phys.\ Rev.\ Lett. {\bf 106}, 259001 (2011), 
  arXiv:1105.5147v2 [astro-ph.HE].
      %
\bibitem{Ch11}
  K.~S.~Cheng, Y.-W.~Yu and T.~Harko, 
  Phys.\ Rev.\ Lett. {\bf 106}, 259002 (2011),
  arXiv:1105.6147v2 [astro-ph.HE].
%
  %
 \bibitem{Planck}
 http://www.esa.int/SPECIALS/Planck/index.html
 %
      %
      %
 \bibitem{Seljak_Pol1}
  U.~Seljak and A.~Slosar, 
  Phys.\ Rev. D {\bf 74}, 063523 (2006),
  arXiv:astro-ph/0604143.
     %
 \bibitem{Pogosian_Pol1}
  L.~Pogosian, I.~Wasserman and M.~Wyman, 
  Phys. Rev. D {\bf 72}, 023513 (2005),
 arXiv:astro-ph/0604141.
   %
 \bibitem{Pogosian_Pol2}
  L.~Pogosian and M.~Wyman, 
  Phys.\ Rev.\ D {\bf 77}, 083509 (2008),
 arXiv:0711.0747v3 [astro-ph].
%
\bibitem{Bevis_Pol1}
   N.~Bevis, M.~Hindmarsh, M.~Kunz and  J.~Urrestilla,
   Phys.\ Rev.\ D {\bf 76}, 043005 (2007),
  arXiv:0704.3800v3 [astro-ph].
   %
\bibitem{Garcia_Pol1}
   J.~Garcia-Bellido, R.~Durrer, E.~Fenu, D.G.~Figueroa and M.~Kunz,
   Phys. Lett. B {\bf 695}, 26 (2011).
  arXiv:1003.0299v2 [astro-ph.CO].
 %
 %
%
%
\bibitem{Vachaspati_GravRad1}
  T.~Vachaspati and A.~Vilenkin, 
 Phys.\ Rev. D {\bf 31}, 3052-3058 (1985).
 %
\bibitem{Hindmarsh_GravRad1}
  M.~Hindmarsh, 
 Phys.\ Lett. {\bf B251}, 1, 28-33 (1990).
  %
\bibitem{Darmour_GravRad1}
  T.~Darmour and A.~Vilenkin,
  Phys.\ Rev.\ Lett. {\bf 85}, 3761-3764 (2000),
  arXiv:gr-qc/0004075.
   %
\bibitem{Darmour_GravRad2}
  T.~Darmour and A.~Vilenkin, 
  Phys.\ Rev. D {\bf 64}, 064008 (2001),
  arXiv:gr-qc/0104026.
     %
\bibitem{Darmour_GravRad3}
  T.~Darmour and A.~Vilenkin,
  Phys.\ Rev. D {\bf 71}, 063510 (2005),
  arXiv:hep-th/0410222.
 %
 %
 %
 %
 \bibitem{LISA/NGO}
 http://lisa.nasa.gov/mission/
 %
 \bibitem{LIGO}
 http://www.ligo.caltech.edu/
 %
   %
\bibitem{Magueijo:1992tt}
  J.~C.~R.~Magueijo,
  Phys.\ Rev.\  D {\bf 46}, 1368-1378 (1992).
%
\bibitem{mag-unpublish}
  J.~C.~R.~Magueijo,
  Report No.\ DAMTP-92-04, 1992 (unpublished).
 %
 %
\bibitem{Parker:1987qx}
  L.~Parker,
  Phys.\ Rev.\ Lett.\  {\bf 59}, 1369 (1987).
  %
\bibitem{Sa88}
  V.~Sahni,
  Mod.\ Phys.\ Lett.\  {\bf 3}, 15, 1425-1429 (1988).
    %
\bibitem{Da88}
  P.~W.~C.~Davies and V.~Sahni,
  Class.\ Quantum\ Grav.\  {\bf 5}, 1-17 (1988).
  %
 %
      %
 \bibitem{Hindmarsh_rev}
  M.~Hindmarsh,
  Lecture Notes in Physics {\bf 360}, 166-171 (1990).
  %
\bibitem{Harada}
T. ~Harada, K.~Nakao and B.~Nolan,
Phys. Rev. D {\bf 80} 024025 (2009), 
Erratum-ibid. D {\bf 80} 109903 (2009), 
arXiv:0812.3462 [gr-qc].
  %
 \bibitem{wald}
 Robert~M.~Wald, {\it General Relativity},
(The University of Chicago Press, Chicago and London, 1984).
 %
 %
 %
  %
   %
    %
\end{thebibliography}
\end{document}